\documentclass[showpacs,showkeys,a4paper,nofootinbib,superscriptaddress]{revtex4-2}
\usepackage{graphicx}
\graphicspath{ {./images/} }

\usepackage{newpxmath,newpxtext}
\usepackage{graphicx}
\usepackage{epsfig}
\usepackage[normalem]{ulem}
\usepackage{xcolor}

\newcommand{\pder}[2]{\dfrac{\partial#1}{\partial#2}}
\newcommand{\pdder}[3]{\dfrac{\partial^2 #1}{\partial #2 \partial #3}}

\newcommand{\dder}[2]{\dfrac{\delta#1}{\delta#2}}
\newcommand{\pdot}[1]{\dot{\partial}_{#1}}

\newcommand{\Gd}{\mathcal{G}}

\newcommand{\R}{\mathcal{R}}
\newcommand{\de}{\mathrm{d}}

\newcommand{\lin}{\\[7pt]}
\usepackage{tensor}

\begin{document}

\title{Raychaudhuri equations, Tidal forces and Weak field Limit in Schwarzshild-Finsler-Randers spacetime.}
\author{A. Triantafyllopoulos}
\email{alktrian@phys.uoa.gr}
\affiliation{Section of Astrophysics, Astronomy and Mechanics, Department of 
Physics, National and Kapodistrian University of Athens, Panepistimiopolis 15784, Athens, Greece}

\author{E. Kapsabelis}
\email{manoliskapsabelis@yahoo.gr}
\affiliation{Section of Astrophysics, Astronomy and Mechanics, Department of 
Physics, National and Kapodistrian University of Athens, Panepistimiopolis 15784, Athens, Greece}

\author{P. C. Stavrinos}
\email{pstavrin@math.uoa.gr}
\affiliation{Department of Mathematics, National and Kapodistrian University of 
Athens,	Panepistimiopolis 15784, Athens, Greece}

\begin{abstract}
In this article, we study the form of deviation of geodesics (tidal forces) and Raychaudhuri equation in a Schwarzshild-Finsler-Randers (SFR) spacetime which has been investigated in previous papers. This model is obtained by considering the structure of a Lorentz tangent bundle of spacetime and in particular the kind of the curvatures in generalized metric spaces  where there is more than one curvature tensor, such as Finsler-like spacetimes. In these cases, the concept of the Raychaudhuri equation is extended with extra terms and degrees of freedom from the dependence of internal variables as the velocity or an anisotropic vector field. {Additionally, we investigate some consequences of the weak field limit on the spacetime under consideration and study the Newtonian limit equations which include a generalization of the Poisson equation.}
\end{abstract}


\maketitle

\section{Introduction}

The evolution equation of the quantities that characterize the (gravitational) flow in a given background spacetime is the Raychaudhuri equation \cite{Raychaudhuri:1953yv,Kar:2006ms}. The flows are integral curves, geodesics or they are generated  by a vector field. Raychaudhuri equation is of significant importance since it describes the dynamical evolution of the gravitational fluid and it is produced by the structure of deviation of nearby geodesics which are dominated by the curvature of space. It is originated by A. Raychaudhuri \cite{Raychaudhuri:1953yv}. When the metric structure of spacetime changes, the equation is modified. The deviation of geodesics and the tidal forces play a fundamental role in general relativity, gravitation and cosmology because of the interaction of curvature of spacetime with matter \cite{Misner:1973prb}. The profound role of the equation of geodesic deviation (EDG) on Riemannian spacetime has been recognized in general relativity for a long time. The observable deviation of two neighboring  geodesics (time-like or null) brings to life an appearance of the curvature of spacetime, namely the detection of curvature expresses a property of the matter sector of spacetime that is given by means of EDG and it is connected with the polarization of gravitational waves and their detection \cite{Hou:2018mey}. How does the small length deviation vector between corresponding points of two nearby geodesics vary as they move along the geodesics? This is the problem of geodesics deviation the solution of which provides a good insight into the nature and behavior  of space. In cosmology, this problem can be connected with the tidal forces and the scale factor $\alpha(t)$ during the expansion of the universe between geodesic motions of two nearby galaxies, see \cite{Hawking:1973uf}. The form of geodesics and their deviation depends on the spacetime metric, the connection and the curvature. Raychaudhuri in his articles \cite{Raychaudhuri:1953yv,Raychaudhuri:1952cec,Raychaudhuri:1953ac, Raychaudhuri:1957rnc} assumes that the Universe can be represented by a time-dependent geometry but does not assume homogeneity or isotropy at early times.  One of his aims is to see whether non-zero rotation (spin), anisotropy (shear) and/or a cosmological constant can succeed in circumventing the initial singularity \cite{Kar:2006ms}. Deviation of geodesics and Raychaudhuri equation can be studied in a more general geometric framework than the Riemannian one. Finsler geometry consists of a natural metric generalization of Riemannian geometry. During the last years, a rapid progress in the field of Finsler geometry and its applications to gravity and cosmology have extended the research in the corresponding topics, we mention some recent works \cite{Kostelecky:2011qz,Caponio:2017lgy,Bubuianu:2018qsq,Pfeifer:2019wus,Javaloyes:2018lex,Javaloyes:2020cc,Hohmann:2020mgs,Caponio:2020ofw,Triantafyllopoulos:2020vkx,Konitopoulos:2021eav,Stavrinos:2021ygh,Hohmann:2021zbt,Javaloyes:2022fmp,Heefer:2022sgt,Bubuianu:2022nef,Savvopoulos:2023qfh,Hama:2022vob,Hama:2023bkl}. In generalized metric spaces such as Finsler or Finsler-like, where the motion/velocity/direction are incorporated in the spacetime structure, internal anisotropy  inherent in the EDG \cite{Asanov1991,Stavrinos1993,balan1999weak} and Raychaudhuri equations is attributed on the framework of the tangent bundle of spacetime manifold, thus extending the concept of volume $\theta$, shear $\sigma$ and vorticity $\omega$ \cite{Stavrinos:2012kv,Stavrinos:2016xyg}. In our theory, the concept of volume $\Theta$ expresses the total volume on the tangent bundle which includes the standard form of volume $\theta$ and the internal anisotropic bulk that is caused by the geometrical structure and its coupling to the standard volume $\theta$ due to the additional degrees of freedom (par. \ref{subsec:raychaudhuri-application}) . Additionally, the form of EDG is  modified with extra terms which are originated  by the connections, torsions and anisotropic curvature tensors. In this  geometric framework, anisotropic tidal phenomena arise  from the internal and external structure of spacetime which are imprinted in the generalized EDG and Raychaudhuri equations. The appearance of extra terms in these equations play the role of additional force fields or self-gravitating actions over spacetime which arise from  the richer geometrical structure.  The concept of non-linear connection in Finsler or Finsler-like spacetime  can be interpreted  as interaction between of external and internal structures of spacetime.  In a Finslerian gravitational theory on the tangent bundle of spacetime curvature effects can be considered as total tidal forces which are produced by the external (horizontal) and internal (vertical) curvature tensors. Different considerations of EDG and Raychaudhuri equations on Finsler and Finsler-like spacetimes have been studied by the one of the authors in \cite{Stavrinos:2016xyg,Stavrinos:2012kv}. Einstein-Finsler-like gravitational  field  equations that govern the motion of matter have been derived in   generalized form of metric spaces on the Lorentz tangent bundle with Finsler-like geometrical structure \cite{Triantafyllopoulos:2020ogl,Triantafyllopoulos:2018bli,Konitopoulos:2021eav}.
These equations have also been given in a different form in \cite{Hawking:1973uf,penrose:1965,hawking:1965,hawking:1966,Wald:1984,Yang:2021fjy,Harko:2012ve,mohajan.rev:2013}.  
In this article, as an additional motivation, we investigate the form of the equation of geodesics  deviation, the Raychaudhuri equation in a Schwarzshild- Finsler-Randers (SFR) space adapted on the Lorentz tangent bundle of  spacetime, thus extending the investigation on the SFR framework we have given in previous works \cite{Triantafyllopoulos:2020vkx,Kapsabelis:2021dpb,Kapsabelis:2022bue}. Some physical consequences are also  given in this article.

This work is organized as follows. In sec. \ref{sec:preliminaries} we present the basic elements of the geometrical structure of the model. In sec. \ref{sec:gen_deviation}, we study and derive the form of deviation of geodesics and paths in completely generalized form, we apply them to the  SFR model and we give some additional information for the deviation equation because of the extra degrees of freedom and the new geometrical concepts. The resulting anisotropic tidal acceleration is of great significance to the investigation of black holes phenomena. Additionally, we give the form of the weak field limit of the deviation equations for the SFR model. In sec. \ref{sec:raychaudhuri1}, we study the generalized Raychaudhuri equations in a general and special form for the model under considaration. We analyze the derived equations in the horizontal and vertical parts of the Lorentz tangent bundle and we give some interpretations to these equations. Finally, in the conclusion (sec. \ref{sec:conclusion}) we discuss and summarize our results.

\section{Geometrical structure of the model}
\label{sec:preliminaries}

In this section, we present some basic elements of the underlying geometry of the SFR gravitational model, as well as the field equations that determine the relation between geometry and matter. A thorough study of this model can be found in \cite{Triantafyllopoulos:2020vkx,Triantafyllopoulos:2020ogl}.

\subsection{The Lorentz tangent bundle}
A Lorentz tangent bundle $TM$ over a spacetime 4-dimensional manifold $M$ is a fibered 8-dimensional manifold with local coordinates $\{x^\mu,y^ a\}$ where the indices of the spacetime variables $x$ are $\kappa,\lambda,\mu,\nu,\ldots = 0,\ldots,3$ and the indices of the fiber variables $y$ are  $ a, b,\ldots,f = 0,\ldots,3$. An extended Lorentzian structure on $TM$
can be provided if the background manifold is equipped with a Lorentz metric tensor
of signature $(-1,\ldots,1)$.

Below, we present some basic geometrical structures of the model.

\subsubsection{The adapted basis}

In order to take a horizontal and vertical basis on a tangent bundle ${TM}$, we need to define a nonlinear connection $\mathbf{N}$ to divide unequivocally the bundle to a horizontal and vertical sub-bundle. The nonlinear connection defines a split of the total space $TTM$ into a horizontal subspace $T_HTM$ and a vertical subspace $T_VTM$. The total space is  the Whitney sum:
    \begin{equation}
        TTM = T_HTM \oplus T_VTM
    \end{equation}
We consider a vector field $X=X^{\mu}\frac{\partial}{\partial x^{\mu}}$ on a base Riemannian manifold ${M}$ along a curve $x^{\mu}(t)$.
If the vector field $X^{\mu}$ coincides with tangent vector of the curve $X^{\mu}=\dot{x}^{\mu}(t)$ then the geodesics equation on ${M}$ is the standard equation using the Levi-Civita connection. It is written as:
\begin{equation}
    \frac{d^{2}x^{\mu}}{ds^{2}} + \Gamma^{\mu}_{\nu\kappa}\dot{x}^{\nu}\dot{x}^{\kappa} = 0
    \label{gr-geodesics}
\end{equation}
The parallel transport of $X$ is given by the equation of geodesics \eqref{gr-geodesics}:
\begin{equation}
\frac{dX^{\mu}}{ds} + \Gamma^{\mu}_{\nu\kappa}X^{\nu}X^{\kappa} = 0  
\label{par-trans}
\end{equation}
where $\Gamma^{\mu}_{\nu\kappa}$ is the metrical connection.

We can extend the vector field $X$ to the tangent bundle ${TM}$ of ${M}$ as
\begin{equation}\label{X-tan-bun}
    \overline{X} = X^{\mu}\frac{\partial}{\partial x^{\mu}} + \frac{dX^{a}}{dt}\frac{\partial}{\partial y^{a}}
\end{equation}
where the coefficients $X^a$ are one by one equal to the coefficients $X^\mu$. Thus, $X^\mu$ and $\frac{dX^{a}}{dt}$ are the coefficients of the extended vector $\overline{X}$ on ${TM}$. the basis for the horizontal and vertical subspace of the tangent bundle. 

According to the considerations of the proposal (4.2) of p.28 of \cite{Miron:1994nvt}
, we can substitute rel.\eqref{par-trans} to rel.\eqref{X-tan-bun} and find:
\begin{equation}
    \overline{X} = X^{\mu}\frac{\partial}{\partial x^{\mu}} - \left(\Gamma^{a}_{\nu\kappa}X^{\nu}X^{\kappa}\right)\frac{\partial}{\partial y^{a}}
\end{equation}
By assuming a Cartan-type connection $\Gamma^{a}_{\mu b}y^{b} = N^{a}_{\mu}$ \cite{miron-watanabe-ikeda:1987} in a Finsler connection, where $N^a_\mu$ are the coefficients of a nonlinear connection, we find:
\begin{equation}
    \overline{X} = X^{\mu}\left(\frac{\partial}{\partial x^{\mu}}-N^{a}_{\mu}\frac{\partial}{\partial y^{a}}\right)=X^{\mu}\delta_{\mu}
\end{equation}
where we defined:
\begin{equation}
    \delta_{\mu}(x,y)=\frac{\partial}{\partial x^{\mu}}-N^{a}_{\mu}\frac{\partial}{\partial y^{a}}
\end{equation}
to be an adapted basis for the tangent bundle. Therefore, the nonlinear connection induces the basis $\{E_A\} = \,\{\delta_\mu,\dot\partial_ a\} $ on the total space, with
\begin{equation}
    \delta_\mu = \dfrac{\delta}{\delta x^\mu}= \pder{}{x^\mu} - N^ a_\mu(x,y)\pder{}{y^ a} \label{delta x}
\end{equation}
and
\begin{equation}
    \dot \partial_ a = \pder{}{y^ a}
\end{equation}

\subsubsection{Metric structure on TM}
A Sasaki-type metric $\Gd$ on $TM$ is:
\begin{equation}
    \mathcal{G} = g_{\mu\nu}(x,y)\,\mathrm{d}x^\mu \otimes \mathrm{d}x^\nu + v_{ a b}(x,y)\,\delta y^ a \otimes \delta y^ b \label{bundle metric}
\end{equation}

A pseudo-Finslerian metric $ f_{ a b}(x,y) $ is defined as one that has a Lorentzian signature of $(-,+,+,+)$ and that also obeys the following form:
\begin{align}
    f_{ a b}(x,y) = \pm\frac{1}{2}\pdder{F^2}{y^ a}{y^ b} \label{Fg}
\end{align}
where the function $F$ satisfies the following conditions \cite{Miron:1994nvt}:
\begin{enumerate}
    \item $F$ is continuous on $TM$ and smooth on  $ \widetilde{TM}\equiv TM\setminus \{0\} $ i.e. the tangent bundle minus the null set $ \{(x,y)\in TM | F(x,y)=0\}$ . \label{finsler field of definition}
    \item $ F $ is positively homogeneous of first degree on its second argument:
    \begin{equation}
        F(x^\mu,ky^ a) = kF(x^\mu,y^ a), \qquad k>0 \label{finsler homogeneity}
    \end{equation}
    \item The form 
    \begin{equation}
        f_{ a b}(x,y) = \pm\dfrac{1}{2}\pdder{F^2}{y^ a}{y^ b} \label{finsler metric} 
    \end{equation}
    defines a non-degenerate matrix: \label{finsler nondegeneracy}
    \begin{equation}
        \det\left[f_{ a b}\right] \neq 0 \label{finsler nondegenerate}
    \end{equation}
\end{enumerate}
where the plus-minus sign in  \eqref{Fg} is chosen so that the metric has the correct signature.

\subsubsection{Connection}
In this work, we consider a distinguished connection ($d-$connection) $ {D} $ on $TM$. This is a linear connection with coefficients $\{\Gamma^A_{BC}\} = \{L^\mu_{\nu\kappa}, L^ a_{ b\kappa}, C^\mu_{\nu c}, C^ a_{ b c} \} $ which preserves by parallelism the horizontal and vertical distributions:
    \begin{align}
        {D_{\delta_\kappa}\delta_\nu = L^\mu_{\nu\kappa}(x,y)\delta_\mu} \quad &,\quad D_{\pdot{ c}}\delta_\nu = C^\mu_{\nu c}(x,y)\delta_\mu \label{D delta nu} \lin
        {D_{\delta_\kappa}\pdot{ b} = L^ a_{ b\kappa}(x,y)\pdot{ a}} \quad &, \quad D_{\pdot{ c}}\pdot{ b} = C^ a_{ b c}(x,y)\pdot{ a} \label{D partial b}
    \end{align}
    From these, the definitions for partial covariant differentiation follow as usual, e.g. for $X \in TTM$ we have the definitions for covariant h-derivative
    \begin{equation}
        X^A_{|\nu} \equiv D_\nu\,X^A \equiv \delta_\nu X^A + L^A_{B\nu}X^B \label{vector h-covariant}
    \end{equation}
    and covariant v-derivative
    \begin{equation}
        X^A|_ b \equiv D_ b\,X^A \equiv \dot{\partial}_ b X^A + C^A_{B b}X^B \label{vector v-covariant}
    \end{equation}
In our consideration, the $d-$connection is metric-compatible:
    \begin{equation}
        D_\kappa\, g_{\mu\nu} = 0, \quad D_\kappa\, v_{ a b} = 0, \quad D_ c\, g_{\mu\nu} = 0, \quad D_ c\, v_{ a b} = 0
    \end{equation}
The $d-$connection coefficients of our model have the following form:
\begin{align}
    L^\mu_{\nu\kappa} & = \frac{1}{2}g^{\mu\rho}\left(\delta_kg_{\rho\nu} + \delta_\nu g_{\rho\kappa} - \delta_\rho g_{\nu\kappa}\right) \label{metric d-connection 1}  \\
    L^ a_{ b\kappa} & = \dot{\partial}_ b N^ a_\kappa + \frac{1}{2}v^{ a c}\left(\delta_\kappa v_{ b c} - v_{ d c}\,\dot{\partial}_ b N^ d_\kappa - v_{ b d}\,\dot{\partial}_ c N^ d_\kappa\right) \label{metric d-connection 2}  \\
    C^\mu_{\nu c} & = \frac{1}{2}g^{\mu\rho}\dot{\partial}_ c g_{\rho\nu} \label{metric d-connection 3} \\
    C^ a_{ b c} & = \frac{1}{2}v^{ a d}\left(\dot{\partial}_ c v_{ d b} + \dot{\partial}_ b v_{ d c} - \dot{\partial}_ d v_{ b c}\right) \label{metric d-connection 4}
\end{align}

\subsubsection{Curvature and torsion}
Curvatures and torsions on $TM$ are defined by the multilinear maps:
    \begin{equation}
        \mathcal{R}(X,Y)Z = [{D}_X,{D}_Y]Z - {D}_{[X,Y]}Z \label{Riemann tensor TM}
    \end{equation}
    and
    \begin{equation}
        \mathcal{T}(X,Y) = {D}_XY - {D}_YX - [X,Y] \label{torsion TM}
    \end{equation}
    where $X,Y,Z \in TTM$.
    We use the following definitions for the curvature components \cite{Miron:1994nvt,Vacaru:2005ht}:
    \begin{align}
        \mathcal{R}(\delta_\lambda,\delta_\kappa)\delta_\nu = R^\mu_{\nu\kappa\lambda}\delta_\mu \label{R curvature components} \lin
        \mathcal{R}(\delta_\lambda,\delta_\kappa)\pdot{ b} = R^ a_{ b\kappa\lambda}\pdot{ a}\lin
        \mathcal{R}(\pdot{ c},\delta_\kappa)\delta_\nu = P^\mu_{\nu\kappa c}\delta_\mu \lin
        \mathcal{R}(\pdot{ c},\delta_\kappa)\pdot{ b} = P^ a_{ b\kappa c}\pdot{ a}\lin
        \mathcal{R}(\pdot{\delta},\pdot{ c})\delta_\nu = S^\mu_{\nu c\delta}\delta_\mu\lin
        \mathcal{R}(\pdot{\delta},\pdot{ c})\pdot{ b} = S^ a_{ b c\delta}\dot{\partial}_ a \label{S curvature components}
    \end{align}
    In addition, we use the following definitions for the torsion components:
    \begin{align}
        \mathcal{T}(\delta_\kappa,\delta_\nu) = & \,\mathcal{T}^\mu_{\nu\kappa}\delta_\mu + \mathcal{T}^ a_{\nu\kappa}\pdot{ a} \label{torsion components 1} \lin
        \mathcal{T}(\pdot{ b},\delta_\nu) = & \,\mathcal{T}^\mu_{\nu b}\delta_\mu + \mathcal{T}^ a_{\nu b}\pdot{ a} \label{torsion components 2} \lin
        \mathcal{T}(\pdot{ c},\pdot{ b}) = & \,\mathcal{T}^\mu_{ b c}\delta_\mu + \mathcal{T}^ a_{ b c}\pdot{ a} \label{torsion components 3}
\end{align}
The h-curvature tensor of the $d-$connection in the adapted basis and the corresponding h-Ricci tensor have, respectively, the components given from \eqref{R curvature components}:
\begin{align}
    & R^\mu_{\nu\kappa\lambda} = \delta_\lambda L^\mu_{\nu\kappa} - \delta_\kappa L^\mu_{\nu\lambda} + L^\rho_{\nu\kappa}L^\mu_{\rho\lambda} - L^\rho_{\nu\lambda}L^\mu_{\rho\kappa} + C^\mu_{\nu a} R^ a_{\kappa\lambda} \label{R coefficients 1}\lin
    & R_{\mu\nu} = R^\kappa_{\mu\nu\kappa} =  \delta_\kappa L^\kappa_{\mu\nu} - \delta_\nu L^\kappa_{\mu\kappa} + L^\rho_{\mu\nu}L^\kappa_{\rho\kappa} - L^\rho_{\mu\kappa}L^\kappa_{\rho\nu}  + C^\kappa_{\mu a} R^ a_{\nu\kappa} \label{d-ricci 1}
\end{align}
where
\begin{equation}\label{Omega}
     R^ a_{\nu\kappa} = \dder{N^ a_\nu}{x^\kappa} - \dder{N^ a_\kappa}{x^\nu}
\end{equation}
are the non-holonomy coefficients, also known as the curvature of the nonlinear connection.

The v-curvature tensor of the $d-$connection in the adapted basis and the corresponding v-Ricci tensor have, respectively, the components \eqref{S curvature components}:
\begin{align}
    S^ a_{ b c d} & = \pdot{d} C^ a_{ b c} - \pdot{ c}C^ a_{ b d} + C^ e_{ b c}C^ a_{ e d} - C^ e_{ b d}C^ a_{ e c} \label{S coefficients 2} \lin
    S_{ a b} & = S^ c_{ a b c} = \pdot{ c}C^ c_{ a b} - \pdot{ b}C^ c_{ a c} + C^ e_{ a b}C^ c_{ e c} - C^ e_{ a c}C^ c_{ e b} \label{d-ricci 4}
\end{align}
The curvature tensor mixed coefficients are:
\begin{align}
		R^a_{b\kappa\lambda} & = \delta_\lambda L^a_{b\kappa} - \delta_\kappa L^a_{b\lambda} + L^c_{b\kappa}L^a_{c\lambda} - L^c_{b\lambda}L^a_{c\kappa} + C^a_{bc} R^c_{\kappa\lambda} \label{R-mixed}\\
		P^\mu_{\nu\kappa c} & = \pdot{c} L^\mu_{\nu\kappa} -  D_\kappa C^\mu_{\nu c} + C^\mu_{\nu b} \mathcal T^b_{\kappa c} \label{Ph} \\
		P^a_{b\kappa c} & = \pdot{c} L^a_{b\kappa} -  D_\kappa C^a_{bc} + C^a_{b d} \mathcal T^d_{\kappa c} \label{Pv}\\
		S^\mu_{\nu c d} & = \pdot{ d} C^\mu_{\nu c} - \pdot{c}C^\mu_{\nu d} + C^\kappa_{\nu c}C^\mu_{\kappa d} - C^\kappa_{\nu d}C^\mu_{\kappa c} \label{S-mixed}
\end{align}
The generalized Ricci scalar curvature in the adapted basis is defined as
\begin{equation}
    \R = g^{\mu\nu}R_{\mu\nu} + v^{ a b}S_{ a b} = R+S \label{bundle ricci curvature}
\end{equation}
where
\begin{align}
    R=g^{\mu\nu}R_{\mu\nu} \quad,\quad
    S=v^{ a b}S_{ a b} \label{hv ricci scalar}
\end{align}

\subsubsection{Hilbert-like action}
A Hilbert-like action on $TM$ can be defined as
    \begin{equation}\label{Hilbert like action}
        K = \int_{\mathcal N} d^8\mathcal U \sqrt{|\Gd|}\, \R + 2\kappa \int_{\mathcal N} d^8\mathcal U \sqrt{|\Gd|}\,\mathcal L_M
    \end{equation}
    for some closed subspace $\mathcal N\subset TM$, where $|\Gd|$ is the absolute value of the metric determinant, $\mathcal L_M$ is the Lagrangian of the matter fields, $\kappa$ is a constant and
    \begin{equation}
        d^8\mathcal U = \de x^0 \wedge \ldots \wedge\de x^3 \wedge \de y^0 \wedge \ldots \wedge \de y^3
    \end{equation}
{ where the 8-parallelepiped  $d^{8}\mathcal{U}$ is considered an oriented compact element of volume.}

\subsection{The SFR model}

In the SFR model, the metric $g_{\mu\nu}$ is the classic Schwarzschild one:
\begin{align}\label{Schwarzchild}
    &g_{\mu\nu}\de x^\mu  \de x^\nu = -fdt^2 + \frac{dr^2}{f} + r^2 d\theta^2 + r^2 \sin^{2}\theta\, d\phi^2
\end{align}
with $f=1-\frac{R_s}{r}$ and $R_s=2GM$ the Schwarzschild radius (we assume units where $c=1$).

The metric $v_{ a b}$ is derived from a metric function $F_v$ of the $ a$-Randers type:
\begin{equation}\label{RandersL}
    F_v = \sqrt{-g_{ a b}(x)y^ a y^ b} + A_ c(x) y^ c
\end{equation}
where $g_{ a b}=g_{\mu\nu}\delta^{\mu}_{ a}\delta^{\nu}_{ b}$ is the Schwarzschild metric and $A_{ c}(x)$ is a covector which expresses a deviation from general relativity, with $|A_ c(x)|\ll 1$. The nonlinear connection will take the form:
\begin{equation}\label{Nconnection}
    N^ a_\mu = \frac{1}{2}y^ b g^{ a c}\partial_\mu g_{ b c}
\end{equation}
The metric tensor $v_{ a b}$ of \eqref{RandersL} is derived from \eqref{Fg} after omitting higher order terms $O(A^2)$:
\begin{equation}\label{vab}
    v_{ a b}(x,y) = g_{ a b}(x) + w_{ a b}(x,y)   
\end{equation}
where 
\begin{align}\label{wab}
    w_{ a b} = & \, \frac{1}{\tilde{a}}(A_{ b}g_{ a c}y^ c + A_{ c}g_{ a b}y^ c + A_{ a}g_{ b c}y^ c) + \frac{1}{\tilde{a}^3}A_{ c}g_{ a e}g_{ b d}y^ c y^ d y^ e
\end{align}
with $\tilde{a} = \sqrt{-g_{ a b}y^{ a}y^{ b}}$.
The total metric defined from the steps above is called the Schwarzschild-Finsler-Randers (SFR) metric and the corresponding spacetime is called an SFR spacetime.

We remark that the term $w_{ab} $ represents a deviation from the pseudo-Riemannian space.
The above-mentioned term can be useful for studying gravitational waves in a locally anisotropic framework of an SFR spacetime.

    Variation of the action \eqref{Hilbert like action} with respect to $g_{\mu\nu}$, $v_{ a b}$ and $N^ a_\kappa$ leads to the following field equations:
\begin{align}
    \overline R_{\mu\nu} - \frac{1}{2}({R}+{S})\,{g_{\mu\nu}}  + \left( \delta^{(\lambda}_\nu \delta^{\kappa)}_\mu - g^{\kappa\lambda}g_{\mu\nu} \right)\left( D_\kappa\mathcal T^ b_{\lambda b} - \mathcal T^ c_{\kappa c}\mathcal T^ b_{\lambda b}\right) & = \kappa {T_{\mu\nu}} \label{feq1}\\
    S_{ a b} - \frac{1}{2}({R}+{S})\,{v_{ a b}}+ \left(v^{ c d}v_{ a b} -  \delta^{( c}_ a \delta^{ d)}_ b \right)\left( D_ c C^\mu_{\mu d} - C^\nu_{\nu c}C^\mu_{\mu d} \right) & = \kappa {Y_{ a b}} \label{feq2}\\
    g^{\mu[\kappa}\pdot{ a}L^{\nu]}_{\mu\nu} +  2 \mathcal T^ b_{\mu b}g^{\mu[\kappa}C^{\lambda]}_{\lambda a} & = \frac{\kappa}{2}\mathcal Z^\kappa_ a \label{feq3}
\end{align}
with
\begin{align}
    T_{\mu\nu} &\equiv - \frac{2}{\sqrt{|\Gd|}}\frac{ \delta\left(\sqrt{|\Gd|}\,\mathcal{L}_M\right)}{ \delta g^{\mu\nu}} = - \frac{2}{\sqrt{-g}}\frac{ \delta\left(\sqrt{-g}\,\mathcal{L}_M\right)}{ \delta g^{\mu\nu}}\label{em1}\\
    Y_{ a b} &\equiv -\frac{2}{\sqrt{|\Gd|}}\frac{ \delta\left(\sqrt{|\Gd|}\,\mathcal{L}_M\right)}{ \delta v^{ a b}}  = -\frac{2}{\sqrt{-v}}\frac{ \delta\left(\sqrt{-v}\,\mathcal{L}_M\right)}{ \delta v^{ a b}}\label{em2}\\
    \mathcal Z^\kappa_ a &\equiv -\frac{2}{\sqrt{|\Gd|}}\frac{ \delta\left(\sqrt{|\Gd|}\,\mathcal{L}_M\right)}{ \delta N^ a_\kappa} = -2\frac{ \delta\mathcal{L}_M}{ \delta N^ a_\kappa}\label{em3}
\end{align}
and $\overline R_{\mu\nu} \equiv R_{\mu\nu} - C^\kappa_{\mu a} R^ a_{\nu\kappa}$, where $\mathcal L_M$ is the Lagrangian of the matter fields, $\delta^\mu_\nu$ and $ \delta^ a_ b$ are the Kronecker symbols, $|\Gd|$ is the absolute value of the determinant of the total metric \eqref{bundle metric}, and
\begin{equation}\label{torsion}
    \mathcal{T}_{\nu b}^{ a} = \pdot{ b} N_{\nu}^{ a} - L_{ b\nu}^{ a}
\end{equation}
are torsion components, where $L_{ b\nu}^{ a}$ is defined in \eqref{metric d-connection 2}. From the form of \eqref{bundle metric} it follows that $\sqrt{|\Gd|} = \sqrt{-g}\sqrt{-v}$, with $g, v$ the determinants of the metrics $g_{\mu\nu}, v_{ a b}$ respectively.

The local anisotropy  can contribute to the energy-momentum tensors of
the horizontal and vertical space $T_{\mu\nu}$ and $Y_{ab}$. As a result, the
energy-momentum tensor $T_{\mu\nu}$ contains the additional information of local
anisotropy of matter fields. $Y_{ab}$, on the other hand, is an extra concept
with no equivalent in Riemannian gravity. It contains more information
about local anisotropy which is produced from the metric $v_{ab}$ which
includes additional internal structure of space-time and can be connected
with the dark energy \cite{Triantafyllopoulos:2018bli}.
  Finally, the energy-momentum tensor $ \mathcal Z^\kappa_ a $ reflects the dependence of
matter fields with respect to the nonlinear connection $N^a_\mu$, a structure
which induces an interaction between internal and external spaces. This
is a different form than of that $T_{\mu\nu}$ and $Y_{ab}$ which depend on just the
external or internal structure respectively.

Solving the field equations \eqref{feq1}-\eqref{feq3} to first order in $A_ c(x)$ in vacuum ($T_{\mu\nu} = Y_{ a b} = \mathcal Z^\kappa_ a = 0$), we get  \cite{Triantafyllopoulos:2020vkx}:
\begin{equation}\label{Asolution}
    A_ c(x) = \left[\tilde A_0 \left|1-\frac{R_S}{r} \right|^{1/2}, 0, 0, 0 \right]
\end{equation}
with $R_S$ the Schwarzschild radius, $r$ the radial coordinate and $\tilde A_0$ a constant, where $|\tilde A_0| \ll 1$.

In the SFR model, the horizontal curvature Ricci tensor $R_{\mu\nu}$ is zero but the internal
vertical curvature v-Ricci tensor $S_{ab}$ and the v-scalar $S$ are different from zero. 
The components of the v-Ricci tensor are:

\begin{align}
	S_{00} & = \frac{\tilde A_0^2 f \left(\tilde a^2-f {y_t}^2\right)^2}{\tilde a^6}  \label{S00} \lin
	S_{11} & = \frac{\tilde A_0^2 \left[{Rs}^2 y_\theta^2 \left(3 \tilde a^2-4 f y_t^2\right)+{Rs}^2 y_\phi^2 \sin ^2\theta  \left(3 \tilde a^2-4 f y_t^2\right)+4 (f-1)^2 f y_t^2 \left(f y_t^2-\tilde a^2\right)\right]}{4 \tilde a^6 (f-1)^2 f} \lin
	S_{22} & = \frac{\tilde A_0^2 {Rs}^2 \left[{Rs}^2 y_\theta^2 \left(4 f y_t^2-3 \tilde a^2\right)-3 \tilde a^2 (f-1)^2 \left(\tilde a^2-f y_t^2\right)\right]}{4 \tilde a^6 (f-1)^4} \lin
	S_{33} & = -\frac{\tilde A_0^2 {Rs}^2 \sin ^2\theta  \left[{Rs}^2 y_\phi^2 \sin ^2\theta  \left(3 \tilde a^2-4 f y_t^2\right)+3 \tilde a^2 (f-1)^2 \left(\tilde a^2-f y_t^2\right)\right]}{4 \tilde a^6 (f-1)^4} \lin
	S_{01} & = -\frac{\tilde A_0^2 {y_t} \left(f y_t^2-\tilde a^2\right) \sqrt{f \left[f y_t^2 - \tilde a^2 - \frac{{Rs}^2 }{(f-1)^2}\left( y_\theta^2 + y_\phi^2 \sin ^2\theta \right)\right]}}{\tilde a^6} \lin
	S_{02} & = \frac{\tilde A_0^2 f {Rs}^2 y_\theta y_t^2 \left(\tilde a^2-f y_t^2\right)}{\tilde a^6 (f-1)^2} \lin
	S_{03} & = \frac{\tilde A_0^2 f {Rs}^2 {y_t} y_\phi\sin ^2\theta  \left(\tilde a^2-f y_t^2\right)}{\tilde a^6 (f-1)^2} \lin
	S_{12} & = \frac{\tilde A_0^2 {Rs}^2 y_\theta{y_r} \left[(f-1)^2 \left(\tilde a^2 f+4 y_r^2\right)+4 f {Rs}^2 y_\theta^2+4 f {Rs}^2 y_\phi^2 \sin ^2\theta \right]}{4 \tilde a^6 (f-1)^4 f^2} \lin
	S_{13} & = \frac{\tilde A_0^2 {Rs}^2 {y_r} y_\phi\sin ^2\theta  \left[(f-1)^2 \left(\tilde a^2 f+4 y_r^2\right)+4 f {Rs}^2 y_\theta^2+4 f {Rs}^2 y_\phi^2 \sin ^2\theta \right]}{4 \tilde a^6 (f-1)^4 f^2} \lin
	S_{23} & = \frac{\tilde A_0^2 {Rs}^4 y_\theta y_\phi\sin ^2\theta  \left(4 f y_t^2-3 \tilde a^2\right)}{4 \tilde a^6 (f-1)^4}
\end{align}

and the scalar v-Ricci curvature is:
\begin{equation}\label{ScurvatureSFR}
    S = \frac{5 \tilde A_0^2 \left[\tilde a^2-f y_t^2\right]}{2 \tilde a^4}
\end{equation}
with $\tilde{a} = \sqrt{-g_{ab}y^{a}y^{b}}$, $ f \equiv 1 - \frac{R_S}{r} $ and we have set $y^0 \equiv y_t, y^1 \equiv y_r, y^2 \equiv y_\theta, y^3 \equiv y_\phi$.

The internal curvature $S^a_{bcd}$ can give a physical meaning to the anisotropy (dependence on direction) of gravitational waves on the SFR model.

We have derived \cite{Kapsabelis:2021dpb} the nontrivial Kretschmann-like invariants of the metrics $g_{\mu\nu}$ and $v_{ab}$ to the lowest non-vanishing order:
\begin{align}
    K_H & \equiv R_{\kappa\lambda\mu\nu}R^{\kappa\lambda\mu\nu} = \frac{12 R_S^2}{r^6} \label{kretschmann h} \\
    K_V & \equiv S_{abcd}S^{abcd} = \left( \frac{3 S}{5} \right)^2 \label{kretschmann v}
\end{align}

Finally, the mixed curvature coefficients are all zero in this model.

\subsection{The Newtonial limit}
{ In this section, we will investigate the Newtonian limit of a Finsler-like metric space on $TM$. The metric on $TM$ will take the form
\begin{equation}
	G = \left(\eta_{\mu\nu} + h_{\mu\nu}(x) \right) \de x^\mu \otimes \de x^\nu + \left( \eta_{ab} + w_{ab}(x,y) \right)\delta y^a \otimes \delta y^\beta
\end{equation}
where $h_{\mu\nu}(x)$ and $w_{ab}(x,y)$ are small perturbations over the flat Minkowski metrics $\eta_{\mu\nu}$ and $\eta_{ab}$ on the horizontal and vertical space respectively.

The field equations \eqref{feq1}, \eqref{feq2} for the metric are written at first order as:
\begin{align}
	R_{\mu\nu} - \frac{1}{2}(R + S) & = \kappa T_{\mu\nu} \label{feqreduced1} \\
	S_{ab} - \frac{1}{2}(R + S) & = \kappa Y_{ab} \label{feqreduced2}
\end{align}
or equivalently
\begin{align}
	\frac{1}{2}\left( \partial_\mu \partial_\kappa h^\kappa_\nu + \partial_\nu\partial_\kappa h^\kappa_\mu - \partial^\kappa\partial_\kappa h_{\mu\nu} - \partial_\mu \partial_\nu h \right) - \frac{1}{2} \eta_{\mu\nu} \left( \partial_\kappa \partial_\lambda h^{\kappa\lambda} - \partial^\kappa \partial_\kappa h + \pdot{a} \pdot{b} w^{ab} - \dot\partial^a \pdot{a} w \right) & = \kappa T_{\mu\nu} \label{feqweak1}\\
	\frac{1}{2}\left( \pdot{c}\pdot{a} w^c_b + \pdot{c} \pdot{b} w^c_a - \dot\partial^c \pdot{c} w_{ab} - \pdot{a}\pdot{b} \right) - \frac{1}{2} \eta_{ab} \left( \partial_\kappa \partial_\lambda h^{\kappa\lambda} - \partial^\kappa \partial_\kappa h + \pdot{a} \pdot{b} w^{ab} - \dot\partial^a \pdot{a} w \right) & = \kappa Y_{ab} \label{feqweak2}
\end{align}
where $h=h^\mu_\mu$ and $w=w^a_a$. The third field equation \eqref{feq3} gives:
\begin{equation}
	\pdot{a}\left( \partial^\kappa h - \partial_\mu h^{\mu\kappa}\right) = \mathcal Z^\kappa_a
\end{equation}
which in our case gives $\mathcal Z^\kappa_a = 0$. This means that the matter fields in our space do not directly depend on the nonlinear connection.
An analytical approach on a weak field metric over a Lorentz tangent bundle can be found in \cite{Triantafyllopoulos:2018bli}.

The horizontal metric is effectively a Riemannian one, so the usual symmetries apply to it. One can decompose this metric on a scalar part $\Phi$, a vector $a_j$, a traceless spatial tensor $b_{ij}$ and the trace $\Psi$ of the spatial part, where all these parts transform independently under spatial rotations. The metric then takes the form:
\begin{align}
	G = & \left\{ -(1 + 2\Phi)\de t^2 + a_i(\de t \de x^i + \de x^i \de t) + \left[ (1-2\Psi)\delta_{ij} + 2 b_{ij}\right] \de x^i \de x^j \right\} \de x^\mu \otimes \de x^\nu  \nonumber\\
	& \, + \left( \eta_{ab} + w_{ab}(x,y) \right)\delta y^a \otimes \delta y^\beta \label{n-limit-metric}
\end{align}
 Additionaly, we can take advantage of the gauge degrees of freedom of the Riemannian metric to set $\partial_i b^{ij} = 0$ and $\partial_i a^i = 0$. Finally, in our setting, the matter content of spacetime is assumed to be dust in its rest frame:
 \begin{equation}
 	T_{\mu\nu} = \rho u_\mu u_\nu
 \end{equation}
where $u^\mu$ is the four velocity field of the matter fluid.
We consider a static spacetime, so all the time derivatives will vanish. 

Under these assumptions, the field equations are written as:
\begin{align}
	\nabla^2\Psi + \frac{1}{4}S & = \frac{\kappa}{2} \rho \label{limit1}\\
	\nabla^2 a_i & = 0 \label{limit2}\\
	(\delta_{ij}\nabla^2 - \partial_i \partial_j)(\Phi - \Psi) - \nabla^2 b_{ij} - \frac{1}{2}S \eta_{ij} & = 0 \label{limit3}
\end{align}
where $\nabla$ is the 3-dimensional spatial grad operator. Taking the trace of \eqref{limit3} yields:
\begin{equation}
	\nabla^2 (\Phi - \Psi) = \frac{3}{4} S \label{limit3trace}
\end{equation}
Substituting \eqref{limit3trace} to \eqref{limit1} gives:
\begin{equation}
	\nabla^2 \Phi = \frac{\kappa}{2} \rho + \frac{S}{2} \label{poisson-like}
\end{equation}
This equation is a direct generalization of the Poisson equation of Newtonian physics. It has been shown in \cite{Triantafyllopoulos:2018bli} that $S$ can describe the effect of a vaccuum energy density, so it can be considered as a dark energy candidate.

Substituting \eqref{poisson-like} to \eqref{limit3} gives:
\begin{equation}
	\frac{1}{4}S \delta_{ij} - \partial_i \partial_j(\Phi - \Psi) - \nabla^2 b_{ij} = 0 \label{metric-tensor}
\end{equation} 
Finally, \eqref{limit2} for a well behaved field gives:
\begin{equation}
	a_i = 0 \label{metric-vector}
\end{equation}
Equations \eqref{poisson-like}, \eqref{metric-tensor} and \eqref{metric-vector} together with \eqref{feqreduced2} (or \eqref{feqweak2}) determine the metric \eqref{n-limit-metric} up to boundary conditions.
The vertical energy-momentum tensor can be approximated by its GR limit value, ie
\begin{equation}
	Y_{ab} = \frac{1}{2}T^\mu_\mu \eta_{ab}
\end{equation}
see also \cite{Triantafyllopoulos:2020vkx}.

We remark that in the Newtonian limit of GR, only the scalar $\Phi$ is nonzero, while in our case more degrees of freedom survive, such as the trace $\Psi$, the traceless tensor $b_{ij}$ and the vertical curvature $S$.}

\section{Generalized deviation of geodesics and paths}
\label{sec:gen_deviation}

In this section, we derive a completely generalized deviation equation using all forms of torsions for the geodesics and the paths. When there are forces acting on particles, they are not moving on geodesics and are accelerating. In our framework, we present some applications to the SFR model.
The deviation equation of geodesics for different types of generalized locally anisotropic spacetime has been studied for a long time \cite{Asanov1991,balan1999weak}. Here, we also present the weak deviation equation for this space.

\subsection{General equations}

We assume that these geodesics and paths take the general form:
\begin{equation}\label{geodesics}
	\frac{dy^a}{d\lambda} + 2G^a = 0 \quad , \quad y^a = \delta^a_\mu \frac{dx^\mu}{d\lambda}
\end{equation}
Geodesics for the SFR model have been derived in a previous work \cite{Kapsabelis:2022bue}:
\begin{equation}\label{SFRgeodesics}
    \ddot{x}^{\lambda} + \Gamma^{\lambda}_{\mu\nu} \dot{x}^{\mu} \dot{x}^{\nu} + g^{\kappa\lambda} \Phi_{\kappa\mu} \dot{x}^{\mu} = 0  
\end{equation}
where $\Gamma^{\lambda}_{\mu\nu}$ are the Christoffel symbols of Riemann geometry, $\dot{x}^{\mu}=\frac{dx^{\mu}}{d\tau}$ and $\Phi_{\kappa\mu}=\partial_{\kappa}A_{\mu}-\partial_{\mu}A_{\kappa}$ and $A_{\mu}$ is the solution rel.\eqref{Asolution}. We notice that from the definition of $\Phi_{\kappa\mu}$ we get a rotation form of geodesics. If $A_{\mu}$ is a gradient of a scalar field, $A_{\mu}=\pder{\Phi}{x^{\mu}}$ then $\Phi_{\kappa\mu}=0$ and the geodesics of our model are identified with the Riemannian ones.

These geodesics are a specific case of \eqref{geodesics} for
\begin{equation}
    G^\lambda = \frac{1}{2} \left( \Gamma^{\lambda}_{\mu\nu} \dot{x}^{\mu} \dot{x}^{\nu} + g^{\kappa\lambda} \Phi_{\kappa\mu} \dot{x}^{\mu} \right)
\end{equation}

The geodesics can be explicitly written in the form:
\begin{align}
	&\ddot t + \frac{1-f}{rf}\dot r \dot t =-\tilde{A}_{0}\dot{r}\frac{f^{-3/2}(1-f)}{2r}
	\label{geodesics0}\\
	&\ddot r + \frac{f(1-f)}{2r} \dot t^2 - \frac{1-f}{2rf} \dot r^2 - rf \big( \dot \theta^2 + \sin^2 \theta \dot  \phi^2 \big)=-\tilde{A}_{0}\dot{t}\frac{f^{1/2}(1-f)}{2r}
	\label{geodesics1}\\[7pt]
	&\ddot \theta + \frac{2}{r} \dot \theta \dot r - \frac{1}{2}\sin 2\theta \, \dot \phi^2=0
	\label{geodesics2}\\
	&\ddot \phi + \frac{2}{r} \dot \phi \dot r + 2\cot \theta \, \dot \theta \dot \phi=0
	\label{geodesics3}
\end{align}
{ The deflection angle of SFR model has been studied in a previous work \cite{Kapsabelis:2022bue} and has been calculated for
the model in hand,
\begin{equation}
	\delta\phi_{SFR}\approx \left(1+\frac{a^{2}}{2} \right)\frac{4GM}{b}
\end{equation}
with $ a = \tilde A_0 b /J$, where $b = J/ER$ is a composite constant formed by the
ratio of the angular momentum $J$ divided by the energy $ER$ of the particle moving along the geodesic.
The deflection angle $\delta\phi$ of GR is $\delta\phi_{SFR}=\delta\phi_{GR}$ when $\lim_{\tilde{A_0}\rightarrow 0}$ .
Connecting the geometrical concept of the curvature $\kappa_{\phi}=\frac{d\phi}{d\tau}$ with the quantity $\kappa_{\phi}=\frac{\delta\phi_{SFR}}{d\tau}$, we get the deflection curvature of the model SFR.
Comparing this result with of Shapiro et al. deflection angle \cite{Shapiro} parameter, we find the value
\begin{equation}
    2 \left(1 + \frac{a^{2}}{2}\right)\simeq 1 + \gamma_{Shap}
\end{equation}
or $a \sim ±0.0141421$ .

The small difference of the deflection angle of the SFR model from the GR one can be
attributed to the Lorentz violations \cite{Kostelecky:2011qz} or on the small amount of energy which is
added to the gravitational potential of SFR model.}

Subsequently, we will calculate the 8-velocity tangent vector to the geodesics \eqref{geodesics} and define a deviation vector of the geodesics. We get:
\begin{equation}
	U = y^\mu\partial_\mu - 2G^a \pdot{a} = y^\mu\delta_\mu + \left( y^\nu N^a_\nu - 2G^a\right) \pdot{a}
\end{equation}
We write the decomposition of $U$ to its horizontal and vertical components as $\{U^A\} = \{u^\mu,v^a\}$.

We define the deviation vector $J$ so that, together with $U$, they form a local coordinate basis:
\begin{equation}
	[U,J] = 0
\end{equation}
After some calculations, the above commutator relation gives
\begin{align}
    U^C  D_C J^B = J^C  D_C U^B + U^A J^C \mathcal T^B_{AC}
\end{align}
Second covariant derivative of the deviation vector is
\begin{align}
	 D^2_U J^I & = U^A  D_A (U^B  D_B J^I)\nonumber\\
	& = U^A  D_A \left(J^B  D_B U^I + U^B J^C \mathcal T^I_{BC}\right) \nonumber\\
	& = \left(J^A  D_A U^B + U^A J^C \mathcal T^B_{AC}\right)  D_B U^I + U^A J^B \big(  [ D_A,  D_B] +  D_B  D_A\big) U^I + U^A  D_A(U^B J^C \mathcal T^I_{BC}) 
 \label{d2j}
\end{align}
We can write
\begin{align}
	U^A  D_A u^\mu & = \frac{du^\mu}{d\lambda} + u^\lambda u^\nu L^\mu_{\nu\lambda} + v^a u^\nu C^\mu_{\nu a}\nonumber\\
	& = u^\lambda u^\nu L^\mu_{\nu\lambda} + v^a u^\nu C^\mu_{\nu a} - 2G^\mu \label{Udk}
\end{align}
where we used \eqref{geodesics}. We define
\begin{equation}\label{Deltamu}
	\Delta^\mu \equiv u^\nu u^\lambda L^\mu_{\nu\lambda} + u^\nu v^a C^\mu_{\nu a} - 2G^\mu
\end{equation}
so we get
\begin{equation}
    U^A  D_A u^\mu = \Delta^\mu
\end{equation}
We can consider $\Delta^\mu$ as a vector expressing the failure of $u^\mu$ to be parallel transported along the geodesic.

Similarly, we find the covariant derivative of the vertical part of $U$:
\begin{align}
	U^A  D_A v^d & = \frac{dv^d}{d\lambda} + u^\lambda v^a L^d_{a\lambda} + v^a v^b C^d_{ab}\nonumber\\
	& =  u^\lambda v^a L^d_{a\lambda} + v^a v^b C^d_{ab}  + \frac{du^\nu}{d\lambda} N^a_\nu + u^\nu \frac{dN^a_\nu}{d\lambda} - 2\frac{dG^a}{d\lambda} \nonumber\\
	& \equiv \Delta^d \label{Deltagamma}
\end{align}
The commutator of the covariant derivatives is given by the following known relation:
\begin{equation}\label{Dcommut}
	[ D_A,  D_B] U^I = \mathcal R^I_{CBA} U^C U^A J^B + \mathcal T^C_{BA}  D_C U^I
\end{equation}
where $\mathcal R^I_{CBA}$ is the generalized curvature tensor of the connection $ D$ on the tangent bundle.
\vspace{5pt}

\noindent
Taking the spacetime part of \eqref{d2j} and after some straightforward calculations, we get the result:
\begin{equation}\label{h-deviation general}
	 D^2_U J^\mu = \big(\mathcal R^\mu_{BCA} +  D_A \mathcal T^\mu_{BC} + \mathcal T^\mu_{BD} \mathcal T^D_{AC}\big) U^A U^B J^C + 2 U^{[A}J^{C]} \mathcal T^\mu_{BC}  D_A U^B + J^B  D_B \Delta^\mu
\end{equation}
where we used relations \eqref{Deltamu} and \eqref{Dcommut}.

Similarly, the fiber part of \eqref{d2j} gives:
\begin{equation}\label{v-deviation general}
	 D^2_U J^d = \big(\mathcal R^d_{BCA} +  D_A \mathcal T^d_{BC} + \mathcal T^d_{BD} \mathcal T^D_{AC}\big) U^A U^B J^C + 2 U^{[A}J^{C]} \mathcal T^d_{BC}  D_A U^B + J^B  D_B \Delta^d
\end{equation}

Equations \eqref{h-deviation general} and \eqref{v-deviation general} denote the deviation equation of horizontal and vertical paths between two
nearby timelike paths on the Lorentzian tangent bundle . If the vectors $\Delta^\mu$ and $\Delta^d$
are equal to zero, the trajectories of nearby observers are geodesics .
Equation \eqref{h-deviation general} is reduced to the standard geodesic deviation equation of general relativity when all the torsion components and the vector $\Delta^\mu$ are equal to zero and in that case, the curvature tensor coincides with the Riemannian one of Levi-Civita connection. The torsion terms in \eqref{h-deviation general} come from the geometry of our space, they play role of a perturbation for the geodesic deviation of GR. By a physical point of view, perturbations of geodesics are affected by extra terms in their equations e.g because of additional mass, gas or dark matter which interact gravitationally during their motion \cite{Savvopoulos:2023qfh}. Tidal acceleration phenomena and anisotropic tidal-field perturbations can appear coming from different sources of spacetime.

\subsection{Application on the weak-field limit}

We investigate first order perturbations of the deviation equation in a weak Finslerian framework on the tangent bundle of the Riemannian space for an SFR space. A weak-field metric takes the form
\begin{equation}
	G = \left[g_{\mu\nu}(x) + h_{\mu\nu}(x,y)\right]\de x^\mu \otimes \de x^\nu + \left[ g_{ab}(x) + w_{ab}(x,y)\right] \delta y^a \otimes \delta y^b
\end{equation}
with $ |h_{\mu\nu}| \ll 1 $ and $ |w_{ab}| \ll 1 $. From relations \eqref{R-mixed}, \eqref{Ph}, \eqref{Pv} and \eqref{S-mixed}, it is straightforward to see that the mixed term curvatures $R^a_{b\kappa\lambda}$, $ P^\mu_{\nu\kappa c} $, $ P^a_{b\kappa c} $ and $ S^\mu_{\nu c d} $ as well as the vertical curvature $ S^a_{bcd} $ are first order on the perturbations $h_{\mu\nu}$ and $w_{ab}$.

At first order on $h_{\mu\nu}$ and $w_{ab}$, the horizontal deviation equation is:
\begin{equation}
	D_U^2 J^\mu = R^\mu_{\kappa\lambda\nu} u^\kappa u^\nu J^\lambda + \tilde H^\mu
\end{equation}
where $\tilde H^\mu$ is a weak correction on the deviation equation, linear on $h_{\mu\nu}$ and $w_{ab}$:
\begin{align}
	\tilde H^\mu = & \,\, \mathcal T^\mu_{\nu b} \mathcal T^b_{\kappa \lambda}u^\kappa u^\nu J^\lambda + P^\mu_{\nu\kappa c} u^\nu v^c J^\kappa + \left( S^\mu_{\nu bc} + D_c \mathcal T^\mu_{\nu b} \right) u^\nu v^c J^b + D_\lambda \mathcal T^\mu_{\nu b} u^\lambda u^\nu J^b \nonumber \\
	& + \mathcal T^\mu_{\nu b} \left[\left(u^\lambda J^b - v^b J^\lambda \right) D_\lambda u^\nu + \left( v^c J^b - v^b J^c \right) D_c u^\nu \right] + J^B D_B \Delta^\mu
\end{align}
The first order vertical deviation equation is:
\begin{align}
    D_U^2 J^a = D_\lambda \mathcal T^a_{\nu\kappa} u^\nu u^\lambda J^\kappa + \mathcal T^a_{\nu\kappa} \left[ \left(u^\lambda J^\kappa - u^\kappa J^\lambda \right) D_\lambda u^\nu + \left( v^c J^\kappa - u^\kappa J^c \right) D_c u^\nu \right] + J^B D_B \Delta^a + \tilde V^a
\end{align}
with
\begin{align}
	\tilde V^a = & \,\, S^a_{b c d} v^b v^d J^c + P^a_{b \kappa c} v^b v^c J^\kappa + \left( R^a_{b \kappa \lambda} + D_b \mathcal T^a_{\lambda\kappa} \right) v^b u^\lambda J^\kappa + D_c \mathcal T^a_{\kappa b} u^\kappa v^c J^b \nonumber \\
	& + \left( D_\lambda \mathcal T^a_{\nu b} + \mathcal T^a_{\nu\kappa} \mathcal T^\kappa_{\lambda b} \right) u^\nu u^\lambda J^b + \mathcal T^a_{\nu b} \mathcal T^b_{\lambda \kappa} u^\nu u^\lambda J^\kappa 
\end{align}
the perturbation on the vertical deviation equation.

We apply the above equations for the SFR model and the geodesics \eqref{SFRgeodesics} and we get:
\begin{equation}\label{SFRdeviation1}
	\nabla^2_U J^\mu = R^\mu_{\nu\kappa\lambda} u^\kappa u^\nu J^\lambda + 2 u^\lambda u^\nu \nabla_\lambda\left( N^a_\nu \pdot{a} J^\mu \right) + J^\lambda g^{\kappa\mu} \, \nabla_\lambda \left( \Phi_{\nu\kappa} u^\nu\right)
\end{equation}
where $\nabla$ is the Levi-Civita connection on the base manifold, $R^\mu_{\nu\kappa\lambda}$ is the Riemann curvature tensor of the classic Schwarzschild spacetiome and we have assumed that the $y-$dependence of $J^\mu$ on $y$ is weak, i.e. $ |\pdot{a}J^\mu| \ll 1 $. Equation \eqref{SFRdeviation1} is the first order generalization of the deviation equation on the SFR model. 

We remark that a variation of $ \Phi_{\nu\kappa} u^\nu $ along the deviation vector $J^\mu$ can induce an anisotropy of $J^\mu$ which varies along the geodesics. In the GR limit, the two last terms in \eqref{SFRdeviation1} vanish and we obtain the classical deviation equation.

The vertical deviation equation in the SFR spacetime is:
\begin{equation}\label{SFRdeviation2}
	D^2_U J^a = J^B D_B \Delta^a
\end{equation}
Relations \eqref{SFRdeviation1}, \eqref{SFRdeviation2} show the rate of change of anisotropic deviation equation (tidal fields) at first order.

\subsection{The Schwarzschild-Finsler-Randers spacetime}

In SFR spacetime, the h-Ricci curvature tensor $R_{\mu\nu}$ and the h-Ricci curvature scalar $R$, defined in \eqref{d-ricci 1} and \eqref{hv ricci scalar} respectively, are both zero. Consequently, the non-zero components of the h-Riemann curvature tensor, defined in \eqref{R coefficients 1}, are equal to the ones of \cite{Misner:1973prb}:
\begin{equation}
    R^{t}{}_{rrt}=2R^{\theta}{}_{r\theta r}=2R^{\phi}{}_{r\phi r}=\frac{R_S}{r^2(R_S-r)}
\end{equation}
\begin{equation}
    2R^{t}{}_{\theta\theta t}=2R^{r}{}_{\theta\theta r}=R^{\phi}{}_{\theta\phi\theta}=\frac{R_S}{r}
\end{equation}
\begin{equation}
    2R^{t}{}_{\phi\phi t}=2R^{r}{}_{\phi\phi r}=-R^{\theta}{}_{\phi\phi\theta}=\frac{R_S \sin^2\theta}{r}
\end{equation}
\begin{equation}
    R^{r}{}_{trt}=-2R^{\theta}{}_{t\theta t}=-2R^{\phi}{}_{t\phi t}= c^2 \frac{R_S (R_S - r)}{r^4}
\end{equation}
The non-zero components of the h-Riemann curvature tensor are the following:

\begin{equation}
    R^{r'}{}_{t'r't'}= -R^{\theta'}{}_{\phi'\theta'\phi'} = -\frac{R_S}{r^3}  
\end{equation}
\begin{equation}
    R^{\theta'}{}_{t'\theta't'}= R^{\phi'}{}_{t'\phi't'} = -R^{r'}{}_{\theta'r'\theta'} = -R^{r'}{}_{\phi'r'\phi'} = \frac{R_S}{2r^3}.  
\end{equation}
 It is fundamental that the geodesic deviation equation shows the tidal acceleration between two observers who are separated by $J^{\mu}$. It takes the form $D^2 J^{\mu'}/D\tau^2 = -R^{\mu'}{}_{t'\nu' t'} J^{\nu'}$. The observable acceleration $(R_S/r^3)c^2 L$ of a body of length $L$ in radial direction extends it and shrinks it in the { lateral} direction by $-(R_S/(2r^3)) c^2 L$. Moving a body to the direction of a black hole causes a spaghetification to the sizes of the body.
 Additionally, the anisotropic curvature $S$ contributes to an anisotropic deformation of the body in our space.

{ The  vertical curvature depends on the position $x$ and the direction $y$, it is related to the intrinsic mechanism of spacetime where the gravitational field is extended  on the total space of the SFR bundle. If we accept that the Schwarzschild spacetime takes an anisotropic structure with a force field (one form) on its metric, the additional energy originates from the internal (vertical) curvature and increases the form of tidal  field  around of a black hole. Consequently, we consider  that both the  horizontal and vertical (anisotropic) curvatures  may affect the radial and lateral motion of an observer.}

\section{Generalized Raychaudhuri equations}
\label{sec:raychaudhuri1}

In this section we investigate the generalized Raychaudhuri equations originated by our model and we give the equations for horizontal and vertical parts of the Lorentz tangent bundle.
\subsection{Horizontal equations}
We define the divergence tensor $B^{\mu}_{\nu}$ as:
\begin{equation}
B^{\mu}_{\nu}=D_{\nu}u^{\mu}\label{hor-div-tensor}
\end{equation}
where $D_{\nu}$ denotes the horizontal covariant derivative and $u^{\mu}$ is a horizontal vector tangent to the geodesic congruence. We calculate the acceleration vector along the direction of $u^{\mu}$ as:
\begin{equation}
\frac{Du^{\mu}}{d\tau}=u^{\nu}D_{\nu}u^{\mu}=u^{\nu}B^{\mu}_{\nu}\label{hor-geo-dev}
\end{equation}
In order to find the deviation of $B^{\mu}_{\nu}$
we can use rel.\eqref{hor-geo-dev}:
\begin{equation}
\frac{DB^{\mu}_{\nu}}{d\tau}=u^{\sigma}D_{\sigma}B^{\mu}_{\nu}=u^{\sigma}D_{\sigma}(D_{\nu}u^{\mu})
\label{hor-dev-div}
\end{equation}
We can use the commutator of $D$ as:
\begin{equation}
[D_{\sigma},D_{\nu}]u^{\mu}= R^{\mu}_{\lambda\sigma\nu}u^{\lambda}-\mathcal{T}^{\lambda}_{\sigma\nu}D_{\lambda}u^{\mu}-R^{a}_{\sigma\nu}D_{a}u^{\mu}
\label{hor-com}
\end{equation}
where $R^{\mu}_{\lambda\sigma\nu}$ is the horizontal curvature tensor, $\mathcal{T}^{\lambda}_{\sigma\nu}$ is the torsion tensor and $R^{a}_{\sigma\nu}$ is the curvature of the non-linear connection.
If we use rels.\eqref{hor-dev-div} and \eqref{hor-com} we have:
\begin{align}
&\frac{DB^{\mu}_{\nu}}{d\tau}= u^{\sigma}[D_{\sigma},D_{\nu}]u^{\mu}+ u^{\sigma}D_{\nu}D_{\sigma}u^{\mu}\Rightarrow\nonumber\\[7pt]
&\frac{DB^{\mu}_{\nu}}{d\tau}=u^{\sigma}\left[R^{\mu}_{\lambda\sigma\nu}u^{\lambda}-\mathcal{T}^{\lambda}_{\sigma\nu}D_{\lambda}u^{\mu}-R^{a}_{\sigma\nu}D_{a}u^{\mu}\right] + \left[D_{\nu}(u^{\sigma}D_{\sigma}u^{\mu})-(D_{\nu}u^{\sigma})(D_{\sigma}u^{\mu})\right]\Rightarrow\nonumber\\[7pt]
&\frac{DB^{\mu}_{\nu}}{d\tau}=R^{\mu}_{\lambda\sigma\nu}u^{\lambda}u^{\sigma}-\mathcal{T}^{\lambda}_{\sigma\nu}u^{\sigma}B^{\mu}_{\lambda}-R^{a}_{\sigma\nu}u^{\sigma}B^{\mu}_{a}-B^{\sigma}_{\nu}B^{\mu}_{\sigma}
\label{hor-ray}
\end{align}
where we have used rel.\eqref{hor-div-tensor} and $u^{\sigma}D_{\sigma}u^{\mu}=0$ since $u^{\mu}$ is tangent to the geodesics.\\[7pt]
We define the horizontal projection tensor as:
\begin{equation}
P^{\mu}_{\nu}=\delta^{\mu}_{\nu}+u^{\mu}u_{\nu} 
\label{hor-proj-tensor}
\end{equation}
The divergence tensor $B^{\mu}_{\nu}$ can be separated in two parts a symmetric and an antisymmetric part. The symmetric can also be separated into a part with trace and a traceless part. We can write the decomposition by using the projection tensor as follows:
\begin{equation}
 B^{\mu}_{\nu}=\frac{1}{3}\theta P^{\mu}_{\nu} + \sigma^{\mu}_{\nu} + \omega^{\mu}_{\nu}
 \label{hor-div-tensor-sep}
\end{equation}
By using the projection tensor in rel.\eqref{hor-ray} we get:
\begin{equation}
P^{\nu}_{\mu}\frac{DB^{\mu}_{\nu}}{d\tau}=-R_{\mu\nu}u^{\mu}u^{\nu}-\mathcal{T}^{\lambda}_{\sigma\mu}u^{\sigma}B^{\mu}_{\lambda}-R^{a}_{\sigma\mu}u^{\sigma}B^{\mu}_{a}-B^{\sigma}_{\mu}B^{\mu}_{\sigma}
\end{equation}
and if we use the decomposition from rel.\eqref{hor-div-tensor-sep} we find:
\begin{equation}  
\frac{d\theta}{d\tau} = -R_{\mu\nu}u^{\mu}u^{\nu}-\mathcal{T}^{\lambda}_{\sigma\mu}B^{\mu}_{\lambda}u^{\sigma}-R^{a}_{\sigma\mu}B^{\mu}_{a}u^{\sigma}-\frac{1}{3}\theta^{2}-\sigma^{\mu\nu}\sigma_{\mu\nu} + \omega^{\mu\nu}\omega_{\mu\nu}
\label{hor-theta}
\end{equation}
The eq.\eqref{hor-theta} is the generalized Raychaudhuri equation for the horizontal space.\\[7pt] As we can see the rel.\eqref{hor-theta} disturbs the rate of the volume  because of the presence of the nonlinear connection $N^{a}_{\mu}$ and the torsion functions  affect the evolution of the gravitational fluid for possible singularities/conjugate points in the universe. In the framework of  a given congruence of timelike geodesics, the expansion $\Theta$ and shear $\sigma_{\mu\nu}$ are described in a generalized form  provide us the generalized type of Raychaudhuri equation \eqref{hor-theta} that  gives additional information on the kinematics. This is possible due to the perturbation of the deviation equation of nearby geodesics or trajectories which was described in paragraph \ref{sec:gen_deviation}. 
\subsection{Vertical equations}
As with the horizontal space we define the divergence tensor as:
\begin{equation}
B^{a}_{b}=D_{b}v^{a}\label{ver-div-tensor}
\end{equation}
where $D_{b}$ denotes the vertical covariant derivative and $v^{a}$ is a vertical vector tangent to the geodesic congruence. 
In order to find the deviation of $B^{a}_{b}$ we have:
\begin{equation}
\frac{DB^{a}_{b}}{d\tau}=v^{c}D_{c}B^{a}_{b}=v^{c}D_{c}(D_{b}v^{a})
\label{ver-dev-div}
\end{equation}\\
We can use the commutator of $D$ as:
\begin{equation}
[D_{c},D_{b}]v^{a}= S^{a}_{d c b}v^{d}-S^{d}_{c b}D_{d}v^{a}
\label{ver-com}
\end{equation}
where $S^{a}_{d c b}$ is the vertical curvature tensor, $S^{d}_{c b}$ is the commutator of the connection coefficients.
If we use rels.\eqref{ver-dev-div} and \eqref{ver-com} we have:
\begin{align}
&\frac{DB^{a}_{b}}{d\tau}= u^{c}[D_{c},D_{b}]v^{a}+ v^{c}D_{b}D_{c}v^{a}\Rightarrow\nonumber\\[7pt]
&\frac{DB^{a}_{b}}{d\tau}=v^{c}\left[S^{a}_{d c b}v^{d}-S^{d}_{c b}D_{d}v^{a}\right] + \left[D_{b}(u^{c}D_{c}v^{a})-(D_{b}v^{c})(D_{c}v^{a})\right]\Rightarrow\nonumber\\[7pt]
&\frac{DB^{a}_{b}}{d\tau}=
S^{a}_{d c b}v^{c}v^{d}-S^{d}_{c b}B^{a}_{d}v^{c}-B^{c}_{b}B^{a}_{c}
\label{ver-ray}
\end{align}
where we have used rel.\eqref{ver-div-tensor} and $v^{c}D_{c}v^{a}=0$ since $v^{a}$ is tangent to the geodesics.\\[7pt]
As with the horizontal part we decompose the divergence tensor as follows:
\begin{equation}
B^{a}_{b}=\frac{1}{3}\tilde{\theta}P^{a}_{b} + \tilde{\sigma}^{a}_{b} + \tilde{\omega}^{a}_{b}
\label{ver-div-tensor-sep}
\end{equation}
where the projection tensor is written as:
\begin{equation}
P^{a}_{b}=\tilde{\delta}^{a}_{b} + v^{a}v_{b}
\label{ver-proj-tensor}
\end{equation}
If we use the projection tensor in eq.\eqref{ver-ray} we get:
\begin{equation}
P^{b}_{a}\frac{DB^{a}_{b}}{d\tau}= -S_{ab}v^{a}v^{b}-S^{d}_{ca}B^{a}_{d}v^{c}-B^{c}_{a}B^{a}_{c}  
\end{equation}
and by using the separation from rel.\eqref{ver-div-tensor-sep} we find:
\begin{equation}
\frac{d\tilde{\theta}}{d\tau} = -S_{ab}v^{a}v^{b}-S^{d}_{ca}B^{a}_{d}v^{c} -\frac{1}{3}\tilde{\theta}^{2}-\tilde{\sigma}^{ab}\tilde{\sigma}_{ab} + \tilde{\omega}^{ab}\tilde{\omega}_{ab}
\label{ver-theta}
\end{equation}
Relation \eqref{ver-theta} is the vertical Raychaudhuri equation. 

\subsection{Application to the SFR model}
\label{subsec:raychaudhuri-application}

The non-holonomy coefficients of the non-linear connection $R^{a}_{\nu\kappa}$ is given by:
\begin{equation}
    R^{a}_{\nu\kappa}=\delta_{\kappa}N^{a}_{\nu}-\delta_{\nu}N^{a}_{\kappa}
\label{curv-N}
\end{equation}
If we use rel.\eqref{delta x} and \eqref{Nconnection} we find:
\begin{align}
&\delta_{\kappa}N^{a}_{\nu}=\partial_{\kappa}N^{a}_{\nu}-N^{e}_{\kappa}\pdot{e}N^{a}_{\nu}\nonumber\\
&\delta_{\kappa}N^{a}_{\nu}=\frac{1}{2}y^{b}\partial_{\kappa}(g^{ac}\partial_{\nu}g_{bc})-\frac{1}{2}y^{f}g^{ed}\partial_{\kappa}g_{df}\pdot{e}(\frac{1}{2}y^{b}g^{ac}\partial_{\nu}g_{bc})\nonumber\\
&\delta_{\kappa}N^{a}_{\nu}=\frac{1}{2}y^{b}\left(\partial_{\kappa}g^{ac}\partial_{\nu}g_{b c} + g^{ac}\partial_{\kappa}\partial_{\nu}g_{b c} + \frac{1}{4}\partial_{\kappa}g_{b c}\partial_{\nu}g^{ac}\right)
\end{align}
So the non-holonomy coefficients of the non-linear connection from \eqref{curv-N} become:
\begin{equation}
R^{a}_{\nu\kappa}=\frac{1}{4}y^{b}\left(\partial_{\kappa}g^{ac}\partial_{\nu}g_{bc}-\partial_{\nu}g^{ac}\partial_{\kappa}g_{bc}\right)
\label{r-non-hol}
\end{equation}
The non holonomy coefficients of the vertical connection $C^{a}_{bc}$ are:
\begin{equation}
    S^{a}_{bc}=C^{a}_{bc}-C^{a}_{cb}=0
\end{equation}
because from rel.\eqref{metric d-connection 4} $C^{a}_{bc}$ is symmetric.\\
The horizontal component of the torsion tensor is given by:
\begin{equation}
    \mathcal{T}^{\lambda}_{\sigma\mu} = L^{\lambda}_{\sigma\mu} - L^{\sigma}_{\mu\sigma} = 0
\end{equation}
because the from rel.\eqref{metric d-connection 1} $L^{\lambda}_{\sigma\mu}$ is symmetric.\\
For simplicity we will take $\sigma^{a}_{b}$ and $\omega^{a}_{b}$ to be zero in both the horizontal and the vertical Raychaudhuri equations. So rel.\eqref{hor-theta} and \eqref{ver-theta} can be written as:
\begin{align}
\label{theta-hor-sfr}
&\frac{d\theta}{d\tau}=-R_{\mu\nu}u^{\mu}u^{\nu} -R^{a}_{\sigma\mu}B^{\mu}_{a}u^{\sigma}-\frac{1}{3}\theta^{2}\\
&\frac{d\tilde\theta}{d\tau}=-S_{ab}v^{a}v^{b}- \frac{1}{3}\tilde\theta^{2}
\label{theta-ver-sfr}
\end{align}
where $R_{\mu\nu}=0$ because it is identified with the GR case in first order of approximation.
Futhermore if we use rel.\eqref{r-non-hol} and take the vertical vector $v^{a}=(-1,0,0,0)$ we find:
\begin{align}
 &\frac{d\theta}{d\tau} = - \frac{1}{4}y^{b}\left(\partial_{\mu}g^{ac}\partial_{\sigma}g_{bc}-\partial_{\sigma}g^{ac}\partial_{\kappa}g_{bc}\right)B^{\mu}_{a}u^{\sigma} - \frac{1}{3}\theta^{2}\\[7pt]
&\frac{d\tilde\theta}{d\tau} = -\frac{2\tilde A_{0}^{2}}{\tilde a^{2}}f \left(1-f\frac{y^{2}_{t}}{\tilde a^{2}} \right)^{2} - \frac{1}{3}\tilde\theta^{2}
\end{align}
Here, we have used that the time-component of the vertical curvature is given by \eqref{S00} which can be interpreted as the evolution of anisotropic expansion $\tilde \theta$.
 By adding equations \eqref{theta-hor-sfr} and \eqref{theta-ver-sfr} we find:
 \begin{align}
 \frac{d\theta}{d\tau} + \frac{d\tilde\theta}{d\tau} + \frac{1}{3}\theta^{2} + \frac{1}{3}\tilde\theta^{2} = -R_{\mu\nu}u^{\mu}u^{\nu} -R^{a}_{\sigma\mu}B^{\mu}_{a}u^{\sigma}-S_{ab}v^{a}v^{b}\Rightarrow\nonumber\\[7pt]
 \frac{d}{d\tau}(\theta+\tilde\theta) + \frac{1}{3}(\theta+\tilde\theta)^{2}-\frac{2}{3}\theta\tilde\theta = -R_{\mu\nu}u^{\mu}u^{\nu} -R^{a}_{\sigma\mu}B^{\mu}_{a}u^{\sigma}-S_{ab}v^{a}v^{b}
 \label{full-ray-sfr}
 \end{align}
 The above mentioned equation \eqref{full-ray-sfr} represents the volumes and their changes, $\theta$ and $\tilde\theta$ denote the standard volume from the horizontal background part of the tangent bundle and the internal anisotropic bulk which is caused by the anisotropic structure. Likewise, $\theta\tilde\theta$ can be considered the coupling of the background volume with its anisotropic bulk during the evolution of world lines and the quantity $\theta + \tilde\theta$ expresses the total volume.

\section{Discussion and conclusion}
\label{sec:conclusion}

The fully developed  equations  that characterize the  flow in a given background spacetime are the Raychaudhuri equations, which are fundamental since they describe the dynamical evolution of the gravitational fluid. They are produced by the structure of deviation of nearby geodesics which is dominated by the curvature of space. In general, the correspondence between fluids and gravity can be represented in a realistic way to understanding current theoretical and observational problems. 

In this article we examine and derive the deviation equation of geodesics and paths as well as the form of Raychaudhuri equation in a completely generalized framework and we apply them to a Schwarschild -Finsler -Randers model in which we showed that the extra terms in \eqref{h-deviation general}, \eqref{v-deviation general}, \eqref{hor-theta}, \eqref{ver-theta}  anisotropically affect the acceleration tidal vector field and the variation of the volume (expansion) during the evolution of fluid lines (geodesics and paths). From a physical point of view, the generalized deviation  geodesics equation is influenced by extra degrees of freedom   e.g because of additional mass, gas, dark matter etc. \cite{Konitopoulos:2021eav,Savvopoulos:2023qfh}, which interact gravitationally during their motion . Tidal acceleration phenomena and anisotropic tidal-field perturbations can appear from different sources of spacetime. It is remarkable  to mention here that acceleration geometrical concepts constitute intrinsic properties on a tangent bundle as e.g, a vector field. The extended geometrical structure of the SFR  model includes the Schwarschild spacetime and gives us additional information on the kinematics because of the extra degrees of freedom reflected in additional terms of torsion, non-linear connection and S-Kretschmann-like curvature invariants  that are imprinted on the corresponding equations. 

Moreover, we investigate the weak field limit of a Finslerian perturbation on a Riemannian spacetime in the cases of deviation equation. {We also study the Newtonian limit of the model and derive a generalized Poisson equation. Additionally, we presented an interesting application for the deviation angle on our generalized framework and compared the result with the one of GR}. We also study the Raychaudhuri equation which is extended in the horizontal and vertical parts of the SFR spacetime. In particular, the concept of non-linear connection in Finsler or Finsler-like spacetime can be geometrically interpreted as interaction between  external and internal structures on the  Lorentz tangent  bundle spacetime. In a more specific case, a physical interpretation of nonlinear connection can relate internal scalar fields with the matter sector of spacetime, for instance in \cite{Konitopoulos:2021eav,Savvopoulos:2023qfh}. It can be understandable from all the derived equations of SFR model that they reduce to standard GR when  all the extra terms of generalized geometrical structure are omitted.  The anisotropic S-curvature  is significant  in our cosmological model since it can provide  addition information for the evolution of gravitational flow lines and the expansion of the universe as well as singularities (focusing/defocusing) of spacetime. This curvature expresses a very small source of anisotropy as it is evident from the equations (60)-(71) in which all the terms are multiplied by  a constant  $\tilde A_{0} \ll 1$ (rel.60), that means all the values of $S$ are very small.  Consequently, from a physical point of view,  it is possible that in a very early period of the universe, the anisotropies of  CMB influence the geometry during cosmological evolution. 

It is of special interest that one investigates the weak field limit in more detail and connect it with the anisotropic polarization of gravitational waves. This research will be a motivation for a future work.

\end{document}